\documentclass[a4paper]{jpconf}
\usepackage{graphicx}
\usepackage{amsmath}
\usepackage{hyperref}
\begin{document}
\title{Lepton Flavor Universality Violation in a 331 Model in \(b\rightarrow s l^+ l^-\) processes}

\author{S Lebbal$^{1,2}$, N Mebarki$^1$ and J Mimouni$^1$}

\address{$^1$ Laboratoire de Physique Mathématique et Physique Subatomique (LPMPS), Université des Frères Mentouti, Constantine 1.\\
 $^2$ Laboratoire de Physique des Rayonnements et de leurs Intéractions avec la Matière (PRIMALAB), Université Hadj Lakhdar, Batna 1.}

\ead{soumia.lebbal@univ-batna.dz}

\begin{abstract}
Lepton Flavor Universality Violation (LFUV) in \(b\rightarrow s l^+ l^-\) processes is investigated in the context of a 331 model. It is shown that in order to explain the experimentally observed deviations from the Standard Model in these FCNC transitions, a non-minimal version of the model has to be considered. We investigate the ability of this model in accomodating the model-independant scenarios currently favored by global fits.
\end{abstract}
\section{Introduction}
Experimental hints of Lepton Flavor Universality Violation (LFUV) have appeared in semi-leptonic B-decays, both in charged and in neutral processes. In fact, disagreements with the SM expectations have shown contributions of non-SM origin of size \(\mathcal{O}(10\%)\) compared to the coresponding SM amplitudes. In particular, four anomalies have appeared in ratios assessing LFU in the decays \(B\longrightarrow D^{(*)}l\bar{\nu}\), \( (l=e,\mu)\) and \(B\longrightarrow K^{(*)}l^{+}l^{-}\). For this latter, the ratio reported by LHCb, which differs from the na\"ive expectation \(R_{K^{(*)}}^{SM}=1\), is \[R_{K^{(*)}}=\frac{\mathcal{B}\left(B\longrightarrow K\mu^{+}\mu^{-} \right)_{(q^{2}\in [q^{2}_{min}, q^{2}_{max}])}}{\mathcal{B}\left(B\longrightarrow K\e^{+}\e^{-} \right)_{(q^{2}\in [q^{2}_{min}, q^{2}_{max}])}}: \left\lbrace \begin{matrix} \begin{aligned}
&R^{exp}_{K[1.1,6.0]}=0.846^{+0.060+0.016}_{-0.054-0.014} \cite{LHCb 2019}, &2.5\sigma\\ 
&R^{exp}_{K^{*}[0.045, 1.1]}=0.66^{+0.11}_{-0.07}\pm 0.03 \cite{LHCb 2017}, &2.7\sigma\\
&R^{exp}_{K^{*}[1.1, 6.0]}=0.69^{+0.11}_{-0.07}\pm 0.03 \cite{LHCb 2017}, &3.0\sigma
\end{aligned}
\end{matrix}
\right. \]
where \(R_{K^{(*)}}\) are measured over specific ranges for the squared dilepton invariant mass \(q^{2}\) (in \(GeV^{2}/c^{4}\)). In the experimental data, the first errors are statistical and the second ones are systematic. A series of theoretical speculations about a possible New Physics (NP) interpretations have emerged, with the hypothesis that the NP responsible for the breaking of LFU is coupled mainly to the third generation of quarks and leptons. A class of particulary motivated models includes those which are based on new (heavy) exotic bosonic mediators at the TeV scale that couple to the leptons and quarks differently. A good candidate of such models is the 331 model, whose heavy (exotic) charged gauge bosons are LFUV processes mediators.\\
The paper is organized as follows: in Sec. 2 we review the main features of the non-minimal 331 model for our specific choice of the parameter \(\beta\). In Sec. 3, we set the problem and describe the general framework of our study. Sec. 4, is dedicated to the neutral gauge boson-mediated NP contribution for the process of interest. In Sec. 5, we present these contributions and compare them to the global analysis scenarios that are currently favord by the data, and finally, in Sec. 6,  we draw our conclusion.
\section{The Model}
We extand the SM by enlarging its gauge group to the broader \(SU(3)_{C}\times SU(3)_{L}\times U(1)_{X}\) group. The miminal construction \cite{minimal construction} is based on placing the left-handed lepton doublets in \(SU(3)_{L}\) triplets that transform in the same way, while the flipped set-up \cite{flipped set-up} is based on perfect quark family replication. In both cases, the gauge bosons couple identically to all lepton and quark families, respectively. Thus, no LFUV can arise from these couplings. In order to generate LFUV from different couplings of the gauge bosons to all fermionic fields, we adopt the non-miminal 331 construction (Model B in ref. \cite{anomaly-free set}). In this construction, the leptons should be grouped in no less than 5 generations (appendix C in ref. \cite{main article}). The electric charge generator in \(SU(3)_{L}\times U(1)_{X}\) is given by
\begin{equation}
Q=T_{3L}+\beta T_{8L}+XI_{3}
\end{equation}
where \(T_{8L}=(1/2\sqrt{3})\)diag\((1,1,-2)\) and \(T_{3L}=(1/2)\)diag\((1,-1,0)\) are the \(SU(3)_{L}\) diagonal generators, \(X\) is the quantum number associated with \(U(1)_{X}\) and \(I_{3}=\) diag\(\left(1, 1, 1 \right)\). The proportionality constant \(\beta\) distiguishes differents 331 models. It is shown that it can have 4 possible values: \(\pm \sqrt{3}, \pm (1/\sqrt{3})\) which  play a key role in determining the electric charges of extra particles and only for some of its values, the gauge bosons have integer charges. For instance, \(\beta=\pm (1/\sqrt{3})\) does not introduce exotic electric charges of fermions in order to cancel the anomalies. Furthermore, for theses value, the heavy gauge bosons have an interger electric charge. The non-minimal set adopted for \(\beta=-1/\sqrt{3}\) in ref. \cite{main article} presents a spectrum that contains 14 charged particles with masses in agreement with the observations (no light particles apart from the SM ones): 9 quarks (6 light SM's and 3 exotic heavy ones) and 5 charged leptons (3 SM's and two exotic).
 \subsection{Fields and representations}
In our case, we adopt the (anomaly free) non-minimal choice for \(\beta=1/\sqrt{3}\) (Model B in ref. \cite{anomaly-free set2}). We introduce the left-handed components together with the right-handed partners of only chaged fermion particles
\begin{enumerate}
\item[(i)] three generations of quarks
\begin{equation}
\begin{aligned}
\label{quark fields}
&Q_{L}^{m}=
\begin{pmatrix}
d_{m}\\u_{m}\\U_{m} 
\end{pmatrix}
; u_{m}^{R}; d_{m}^{R}; U_{m}^{R}, &\left( m=1,2\right)
\end{aligned}
\end{equation}
with \(SU(3)_{C}\times SU(3)_{L}\times U_{X}(1)\) quantum numbers \(\left( 3, \bar{3}, \frac{1}{3} \right), \left( \bar{3}, 1, 2/3\right), \left( \bar{3},1,-1/3\right), \left( \bar{3}, 1, 2/3\right)\).
\begin{equation}
\begin{aligned}
\label{quark fields}
&Q_{L}^{3}=
\begin{pmatrix}
u_{3}\\u_{3}\\D_{3} 
\end{pmatrix}
; u_{3}^{R}; d_{3}^{R}; D_{3}^{R}
\end{aligned}
\end{equation}
with \(SU(3)_{C}\times SU(3)_{L}\times U_{X}(1)\) quantum numbers \(\left( 3, 3, 0 \right), \left( \bar{3}, 1, 2/3\right), \left( \bar{3},1,-1/3\right), \left( \bar{3}, 1, -1/3\right)\).
\item[(ii)] five generations of leptons
\begin{equation}
\begin{aligned}
\label{lepton fields}
l_{1}^{L}=
\begin{pmatrix}
e^{-L}_{1}\\\nu^{L}_{1}\\N^{L}_{1}
\end{pmatrix}
;&e_{1}^{-R}&
\end{aligned}
\end{equation}
with \(SU(3)_{C}\times SU(3)_{L}\times U_{X}(1)\) quantum numbers \(\left( 1, \bar{3}, -1/3 \right), \left( 1, 1, -1\right)\).
\begin{equation}
\begin{aligned}
\label{lepton fields}
l_{n}^{L}=
\begin{pmatrix}
\nu^{L}_{n}\\e^{-L}_{n}\\E^{-L}_{n}
\end{pmatrix}
;&e_{n}^{-R}; E_{n}^{-R}, &\left( n=2,3\right)
\end{aligned}
\end{equation}
with \(SU(3)_{C}\times SU(3)_{L}\times U_{X}(1)\) quantum numbers \(\left( 1, 3, -2/3 \right), \left( 1, 1, -1\right), \left( 1, 1, -1\right)\).
\begin{equation}
\begin{aligned}
\label{lepton fields}
l_{4}^{L}=
\begin{pmatrix}
N^{L}_{4}\\E^{-L}_{4}\\F^{-L}_{4}
\end{pmatrix}
;&E_{4}^{-R}; F_{4}^{-R}&
\end{aligned}
\end{equation}
with \(SU(3)_{C}\times SU(3)_{L}\times U_{X}(1)\) quantum numbers \(\left( 1, 3, -2/3 \right), \left( 1, 1, -1 \right), \left( 1, 1, -1\right)\).
\begin{equation}
\begin{aligned}
\label{lepton fields}
l_{5}^{L}=
\begin{pmatrix}
\left( E^{-R}_{4}\right) ^{c}\\N^{L}_{5}\\N^{L}_{6}
\end{pmatrix}
\end{aligned}
\end{equation}
with \(SU(3)_{C}\times SU(3)_{L}\times U_{X}(1)\) quantum numbers \(\left( 1, 3, 1/3 \right)\).
\end{enumerate}
It should be stressed that, originally, the charged lepton of the fifth generation should be positive (\(E^{+L}_{5}\), with a right handed component \(E^{+R}_{5}\)); but after the \(\Lambda_{NP}\) scale SSB, this exotic degree of freedom will be masseless (together with \(E_{4}^{-}\)). Meaning that, at the \(EW\) scale, not only the 3 SM charged leptons would acquire mass, but also the two exotic ones, and because the spectrum should contain no light particles apart from the SM ones, we have to get rid of such presence. To do so, we identify the \(E^{+L}_{5}\) with the charge conjugate of the right handed component of \(E^{-}_{4}\). Thus, the right handed part of \(E^{-}_{4}\) should belong to the lepton triplet \(l_{5}^{L}\) rather than being a singlet, besides, there would be no \(E^{+R}_{5}\).
The model contains 16 charged fermions: 9 quarks (3 lignt SM's plus 3 heavy exotic ones) and 7 leptons (3 SM ones plus 4 exotic). The \(SU(3)_{L}\) gauge bosons are denoted in the matrix form by \(W_{\mu}=W_{\mu}^{a}T^{a}\), where \(T^{a}=\lambda^{a}/2\) are the generators of \(SU(3)_{L}\) (\(\lambda^{a}\) being the Gell-Mann matrices and \(a=1,..8)\). For our specific choice of \(\beta\), the neutral ones are \(W_{\mu}^{3}\), \(W_{\mu}^{8}\), \(W_{\mu}^{6}\) and \(W_{\mu}^{7}\). In what follows, we introduce the flavor vectors where the fields (relevant for our process) with the same electric charge are gathered.
\begin{equation}
\begin{aligned}
&D=\left( d_{1},d_{2},d_{3},D_{3}\right)^{T} &\\
&f^{-}=\left( e^{-}_{1},e^{-}_{2},e^{-}_{3},E^{-}_{2},E^{-}_{3},E^{-}_{4},F^{-}_{4}\right)^{T} &
\end{aligned}
\end{equation}
\subsection{Symmetry breaking and spectrum}
The model undergoes two stages of Spontaneous Symmetry Breakings (SSB). The first, triggered by a sextet and a triplet \cite{scalar sector}, occurs at an energy scale \(\Lambda_{NP}\sim TeV\) and allows one to recover the SM. At the order of \(\Lambda_{NP}\), all exotic particles acquire mass: 7 fermions (4 leptons ans 3 quarks), and 5 quage bosons \(W_{\mu}^{4}\), \(W_{\mu}^{5}\), \(W_{\mu}^{6}\) and \(W_{\mu}^{7}\). The two neutal \(X_{\mu}\) and \(W_{\mu}^{8}\) yield a massive \(Z_{\mu}^{'}\) and a massesless one \(B_{\mu}\), with a mixing angle \(\theta_{331}\)
\begin{equation}
\begin{pmatrix}
Z_{\mu}^{'}\\B_{\mu}
\end{pmatrix}
=
\begin{pmatrix}
\cos\theta_{331}& \sin\theta_{331}\\-\sin\theta_{331}& \cos\theta_{331}
\end{pmatrix}
\begin{pmatrix}
X_{\mu}\\W^{8}_{\mu}
\end{pmatrix}
\end{equation} 
Where 
\begin{equation}
\begin{aligned}
&\sin\theta_{331}=\frac{g}{\sqrt{g^{2}+\frac{g^{2}_{X}}{18}}}, &\cos\theta_{331}=\frac{\frac{g_{X}}{3\sqrt{2}}}{\sqrt{g^{2}+\frac{g^{2}_{X}}{18}}}&
\end{aligned}
\end{equation}
Here, \(g\) and \(g_{X} \) are the gauge coupling constants. The subsequent one occurs at energy scale \(\Lambda_{EW}\). It reproduces the EWSB of the SM and is accomplished by means of two triplets and two sextets. At this stage, the remaining SM fermions (except for the neutrinos) and the 3 gauge fields should all acquire a mass. The neural bosons \(W_{\mu}^{3}\) and \(B_{\mu}\) mix together with a mixing angle \(\theta_{W}\) (Weinberg angle) to yield  a massesless \(A_{\mu}\) identified with the photon and a massive \(Z_{\mu}\)  (at the order of interest \(\mathcal{O}(\epsilon^{2})\))
\begin{equation}
\begin{pmatrix}
Z_{\mu}\\A_{\mu}
\end{pmatrix}
\sim
\begin{pmatrix}
\cos\theta_{W}& -\sin\theta_{W}\\ \sin\theta_{W}& \cos\theta_{W}
\end{pmatrix}
\begin{pmatrix}
W^{3}_{\mu}\\B_{\mu}
\end{pmatrix}
\end{equation} 
where, the two mixing angles \(\theta_{331}\) and \(\theta_{W}\) and the two gauge coupling constants obey the relations
\begin{equation}
\label{relation between two gauge coupling constants}
\begin{aligned}
&\cos\theta_{331}=\frac{1}{\sqrt{3}}\tan\theta_{W},&  \frac{g}{g_{X}}=\frac{\tan\theta_{331}}{3\sqrt{2}}&
\end{aligned}
\end{equation}
Before proceeding, we should mention that the neutral leptons are left out of the discussion.
\section{General Framework} 
The total effective Hamiltonian, at the \(b-\)mass scale, for the quark-level transition \(b\rightarrow s l_i^+ l_j^-\) in the presence of NP operators is expressed as
\begin{equation}
\label{SM Hamiltonian}
\mathcal{H}_{eff}\left(b\rightarrow s l_i^+ l_j^-\right)=\mathcal{H}^{SM}_{eff}+\mathcal{H}^{NP}_{eff}=-\frac{4G_{F}}{\sqrt{2}}V^{*}_{ts}V_{tb} \sum_{i} C_i\mathcal O_i
\end{equation}
where
\begin{equation}
\label{total Hamiltonian}
\begin{aligned}
\mathcal{H}^{SM}_{eff}=-\frac{4G_{F}}{\sqrt{2}}V^{*}_{ts}V_{tb} \left\lbrace  \sum_{i=1}^{6} C_i(\mu)\mathcal O_i(\mu)+C_7\frac{e}{16\pi^2}\left[ \bar{s}\sigma_{\mu\nu}(m_sP_L+m_bP_R)b\right] F^{\mu\nu} \right. \\
 \left. +C_{9}\frac{\alpha}{4\pi}\left( \bar{s}\gamma^{\mu}P_{L}b\right) \left( \bar{l_{i}}\gamma_{\mu}l_{j}\right) +C_{10}\frac{\alpha}{4\pi}\left( \bar{s}\gamma^{\mu}P_{L}b\right) \left( \bar{l_{i}}\gamma_{\mu}\gamma_{5}l_{j}\right) \right\rbrace 
\end{aligned}
\end{equation}
Here, \(P_{L,R}=\left( 1\mp\gamma_{5}\right) /2\) and \(\alpha=e^{2}/4\pi\) is the fine-structure constant. The six-dimentional operators \(\mathcal O_i\left( i=1,..6\right) \) correspond to \(P_{i}\) in ref. \cite{Pi operators} and \(C_i\) are the Wilson coefficients. In the SM, only \(\mathcal{O}_{7} \), \(\mathcal{O}_{9} \) and \(\mathcal{O}_{10} \) are significant at the scale \(\mu=m_{b}\). As the analyses of the \(b\rightarrow s\) transitions indicate that the observed pattern of deviations is consistent with a larger vector/axial (VA) contribution \(\left( C^{\mu}_{9}, C^{\mu}_{10}\right) \), we will focus only on the assumed larger VA contributions. The Hamiltonian for the NP contribution related to the deviations seen in the transition \(b\rightarrow s l_i^+ l_j^-\) is thus
\begin{equation}
\begin{aligned}
\label{VA contribution}
\mathcal{H}^{NP}_{eff}=\mathcal{H}^{VA}_{eff}= -\frac{4G_{F}}{\sqrt{2}}V^{*}_{ts}V_{tb}\frac{\alpha}{4\pi} \left\lbrace \left[R_{V}\left( \bar{s}\gamma^{\mu}P_{L}b\right) \left( \bar{l_{i}}\gamma_{\mu}l_{j}\right) +R_{A}\left( \bar{s}\gamma^{\mu}P_{L}b\right) \left( \bar{l_{i}}\gamma_{\mu}\gamma_{5}l_{j}\right)\right] \right. \\ 
\left. +\left[ R^{'}_{V}\left( \bar{s}\gamma^{\mu}P_{R}b\right) \left( \bar{l_{i}}\gamma_{\mu}l_{j}\right) +R^{'}_{A}\left( \bar{s}\gamma^{\mu}P_{R}b\right) \left( \bar{l_{i}}\gamma_{\mu}\gamma_{5}l_{j}\right)\right] \right\rbrace 
\end{aligned}
\end{equation}
where \(R_{V,A}\) are NP effective couplings. The VA contribution can only come from the neutral gauge bosons \(Z_{\mu}^{'}\),\(Z_{\mu}\), \(A_{\mu}\) and \(W_{\mu}^{6,7}\).
\section{Neutral gauge bosons contributions}
In what follows, we will consider only the contributions at the lowest order in \(\epsilon\) where \( \left( \epsilon=\Lambda_{EW}/\Lambda_{NP}\right)\), and we will focus only on the non-SM contributions to the Wilson Coefficients of interest \( R^{\left( '\right) }_{V}\equiv C_{9}^{\left( '\right) }\) and \(  R^{\left( '\right) }_{A}\equiv C_{10}^{\left( '\right) }\) in eq. \eqref{VA contribution}. We define the unitary rotation matrices relating (unprimed) fermion interaction eigenstates and (primed) mass eigenstates
\begin{equation}
\begin{aligned}
&f^{L}=V^{(f)}f^{'L}, &f^{R}=W^{(f)}f^{'R}
\end{aligned}
\end{equation} 
When diagonalizing the mass matrix using perturbation theory in powers of \(\epsilon\) we get:
\begin{itemize}
\item[(i)] at order \(\epsilon^{0}\), each of the masseless SM particles and heavy exotic fermions mix only among themselves.
\item[(ii)] at order \(\epsilon^{1}\), there is only mixing between SM and exotic particles.
\item[(iii)] at order \(\epsilon^{2}\), there is mixing among all the particles of the same electric charge.
\end{itemize}
\subsection{Neutral currents mediated by \(Z_{\mu}\) and \(Z^{'}_{\mu}\) bosons }
The part of the interaction Lagrangian density that gives the couplings of fermions to neutral gauge boson \(Z^{'}_{\mu}\) which is relevant to the process is (in the interaction eigenbasis)
\begin{equation}
\label{lagZ'}
\begin{aligned}
&\mathcal{L}_{Z^{'}} \supset \dfrac{\cos\theta_{331}}{g_{x}}Z^{'}_{\mu} \left\lbrace \bar{D}^{L}\gamma_{\mu}
\begin{pmatrix}
\frac{9g^{2}+g_{x}^{2}}{3\sqrt{6}} & 0 & 0 & 0 \\
0 & \frac{9g^{2}+g_{x}^{2}}{3\sqrt{6}} & 0 & 0 \\
0 & 0 & -\sqrt{\frac{3}{2}}g^{2} & 0 \\
0 & 0 & 0 & \sqrt{6}g^{2} \\
\end{pmatrix}
D_{L}
-\frac{g_{x}^{2}}{3\sqrt{6}}\bar{D}_{R}\gamma_{\mu}D_{R} \right.&\\
& \left. +\bar{f^{-}_{L}}\gamma_{\mu}
\begin{pmatrix}
\frac{9g^{2}-g_{x}^{2}}{3\sqrt{6}} & 0 & 0 & 0& 0 & 0 & 0 \\
0 & \frac{-9g^{2}-g_{x}^{2}}{3\sqrt{6}} & 0 & 0 & 0 & 0 & 0 \\
0 & 0 & \frac{-9g^{2}-g_{x}^{2}}{3\sqrt{6}} & 0 & 0 & 0 & 0 \\
0 & 0 & 0 & \frac{18g^{2}-2g_{x}^{2}}{3\sqrt{6}}  & 0 & 0 & 0 \\
0 & 0 & 0 & 0 &  \frac{18g^{2}-2g_{x}^{2}}{3\sqrt{6}}  & 0 & 0 \\
0 & 0 & 0 & 0 & 0 & \frac{-9g^{2}-g_{x}^{2}}{3\sqrt{6}} & 0 \\
0 & 0 & 0 & 0 & 0 & 0 & \frac{18g^{2}-2g_{x}^{2}}{3\sqrt{6}} \\
\end{pmatrix}
f^{-}_{L} \right. &\\
 &\left. +\bar{f^{-}_{R}}\gamma_{\mu}
\begin{pmatrix}
\frac{-g_{x}^{2}}{\sqrt{6}} & 0 & 0 & 0& 0 & 0 & 0 \\
0 & \frac{-g_{x}^{2}}{\sqrt{6}} & 0 & 0 & 0 & 0 & 0 \\
0 & 0 & \frac{-g_{x}^{2}}{\sqrt{6}} & 0 & 0 & 0 & 0 \\
0 & 0 & 0 & \frac{-g_{x}^{2}}{\sqrt{6}}  & 0 & 0 & 0 \\
0 & 0 & 0 & 0 &  \frac{-g_{x}^{2}}{\sqrt{6}} & 0 & 0 \\
0 & 0 & 0 & 0 & 0 & \frac{-9g^{2}+g_{x}^{2}}{3\sqrt{6}} & 0 \\
0 & 0 & 0 & 0 & 0 & 0 & \frac{-g_{x}^{2}}{\sqrt{6}} \\
\end{pmatrix}
f^{-}_{R} \right\rbrace &
\end{aligned}
\end{equation}
It is clear that the restriction of the interaction matrix \(\mathcal{L}_{Z^{'}}\) to the SM particles is not proportional to the identity matrix in flavor space. So, the FCNC transition arises already before moving to the mass eigenbasis. Due to the \(\mathcal{O}(\epsilon^{2})\) suppresion, compared to the SM, that results from the heavy mass of the \(Z_{\mu}^{'}\) boson, we conclude that the NP contribution from the \(Z_{\mu}^{'}\) boson starts at \(\mathcal{O}(\epsilon^{2})\).\\
For the interaction with \(Z_{\mu}\)
\begin{equation}
\label{lagZ}
\begin{aligned}
&\mathcal{L}_{Z} \supset\cos\theta_{W}gZ_{\mu}\left\lbrace \bar{D}_{L}\gamma_{\mu}
\begin{pmatrix}
-\frac{1+\cos^{2}\theta_{331}}{2} & 0 & 0 & 0 \\
0 & -\frac{1+\cos^{2}\theta_{331}}{2} & 0 & 0  \\
0 & 0 & -\frac{1+\cos^{2}\theta_{331}}{2} & 0 \\
0 & 0 & 0 & \cos^{2}\theta_{331} \\
\end{pmatrix}
D_{L}
+\cos^{2}\theta_{331}\bar{D}_{R}\gamma_{\mu}D_{R} \right. &\\
&\left. +\bar{f^{-}_{L}}\gamma_{\mu}
\begin{pmatrix}
\frac{-1+3\cos^{2}\theta_{331}}{2}& 0 & 0 & 0& 0 & 0 & 0 \\
0 & \frac{-1+3\cos^{2}\theta_{331}}{2} & 0 & 0 & 0 & 0 & 0 \\
0 & 0 & \frac{-1+3\cos^{2}\theta_{331}}{2} & 0 & 0 & 0 & 0 \\
0 & 0 & 0 & 3\cos^{2}\theta_{331}  & 0 & 0 & 0 \\
0 & 0 & 0 & 0 &  3\cos^{2}\theta_{331}  & 0 & 0 \\
0 & 0 & 0 & 0 & 0 & \frac{-1+3\cos^{2}\theta_{331}}{2} & 0 \\
0 & 0 & 0 & 0 & 0 & 0 & 3\cos^{2}\theta_{331}  \\
\end{pmatrix}
f^{-}_{L} \right.&\\
&\left. +\bar{f^{-}_{R}}\gamma_{\mu}
\begin{pmatrix}
3\cos^{2}\theta_{331}& 0 & 0 & 0& 0 & 0 & 0 \\
0 & 3\cos^{2}\theta_{331} & 0 & 0 & 0 & 0 & 0 \\
0 & 0 & 3\cos^{2}\theta_{331} & 0 & 0 & 0 & 0 \\
0 & 0 & 0 & 3\cos^{2}\theta_{331}  & 0 & 0 & 0 \\
0 & 0 & 0 & 0 &  3\cos^{2}\theta_{331} & 0 & 0 \\
0 & 0 & 0 & 0 & 0 & \frac{1-3\cos^{2}\theta_{331}}{3\sqrt{6}} & 0 \\
0 & 0 & 0 & 0 & 0 & 0 & 3\cos^{2}\theta_{331} \\
\end{pmatrix}
f^{-}_{R}\right\rbrace &
\end{aligned}
\end{equation}
In the case of the SM gauge boson, The \(b\rightarrow s\) transition arises at \(\mathcal{O}(\epsilon^{2})\), and because there is no \(\mathcal{O}(\epsilon^{2})\) suppression due to the bososn mass, the NP contribution for the \({Z_{\mu}}\)  starts also at \(\mathcal{O}(\epsilon^{2})\). The interactions of the right hadded quarks, with both neutral bosons, are proportional to the identity in flavor space, so no flavor change can arise at any order in \(\epsilon\) . We conclude that \(Z_{\mu}^{'}\) and \(Z_{\mu}\) do not contribute to \(C_{9}^{'}\) and \(C_{10}^{'}\). Our model does not allow for any contrinution to \(C^{'}_{9,10}\) in the process.\\
\subsection{Neutral currents mediated by \(A_{\mu}\) and \(W^{6,7}_{\mu}\)}
For the photon \(A_{\mu}\), the interaction with down-type quarks is proportional to the identity matrix in the flavor space
\begin{equation}
\mathcal{L_{A}}=\sqrt{3}\cos\theta_{331}\cos\theta_{W}g A_{\mu}\left[ \frac{1}{3}\bar{D}\gamma^{\mu}D\right] 
\end{equation} 
So, there are no FCNC from the photon interaction. As for \(W_{\mu}^{6,7}\), their contributions \cite{big article} to the process are of the \(\mathcal{O}(\epsilon^{3})\). So, they can be neglected compared to \(Z_{\mu}^{'}\) and \(Z_{\mu}\)'s.
\subsection{The NP contribution}
Moving to the mass eigenbasis, and exploiting the unitarity of the rotation matrices \(V\) and \(W\) for the \(\delta_{ij}\) contribution (\(i\) and \(j\) refer to the SM lepton generations), we eliminate the coupling \(g\) by means of eq. \eqref{relation between two gauge coupling constants}, the leading-order \(Z_{\mu}^{'}\) and \(Z_{\mu}\) contributions in terms of effective operators read 
\begin{equation}
\begin{aligned}
\mathcal{H}^{Z^{'}}_{eff}\supset\frac{g_{x}^{2}}{108\cos^{2}\theta_{331}}\frac{1}{M_{Z^{'}}^{2}}V_{3k}^{*}V_{3l}\frac{4\pi}{\alpha}
\left\lbrace \left[ \left( \frac{1+9\cos^{2}\theta_{331}}{2}\right)  \delta_{ij}-V_{1i}^{*}V_{1j}\right] \mathcal{O}_{9}^{ijkl} \right.\\
\left. +\left[ \left( \frac{3\cos^{2}\theta_{331}-1}{2}\right) \delta_{ij}+V_{1i}^{*}V_{1j}\right]  \mathcal{O}_{10}^{ijkl}\right\rbrace 
\end{aligned}
\end{equation}
and
\begin{equation}
\begin{aligned}
&\mathcal{H}^{Z}_{eff}\supset\frac{\cos^{2}\theta_{W}\left( 1+3\cos^{2}\theta_{331}\right) }{8}\frac{g^{2}}{M_{Z}^{2}}\frac{4\pi}{\alpha} V_{4k}^{*}V_{4l} \delta_{ij} \left[ \left( -1+9\cos^{2}\theta_{331}\right) \mathcal{O}_{9}^{ijkl}+\left( 1+3\cos^{2}\theta_{331}\right) \mathcal{O}_{10}^{ijkl}\right] &
\end{aligned}
\end{equation}
where the indices \(k\), \(l\) refer to the SM generations of the quark mass eigenstates (assuming \(k\neq l\)). The NP contributions can be represented by the two quanties \(f_{Z^{'}}\) and \(f_{Z}\) where
\begin{equation}
\begin{aligned}
&\frac{g_{x}^{2}}{108\cos^{2}\theta_{331}}\frac{1}{M_{Z^{'}}^{2}}V_{3k}^{*}V_{3l}\frac{4\pi}{\alpha}=f_{Z^{'}}\left( -\frac{4G_{F}}{\sqrt{2}}V^{*}_{ts}V_{tb}\right)&\\
&\Rightarrow f_{Z^{'}}=-\frac{1}{2\sqrt{2}G_{F}V_{tb}V^{*}_{ts}}\frac{4\pi}{\alpha}\frac{1}{6-2\tan^{2}\theta_{W}}\frac{g^{2}}{M_{Z^{'}}^{2}}V^{*}_{3k}V_{3l}&
\end{aligned} 
\end{equation}
and
\begin{equation}
\begin{aligned}
&\frac{\cos^{2}\theta_{W}\left( 1+3\cos^{2}\theta_{331}\right) }{8}\frac{g^{2}}{M_{Z}^{2}}\frac{4\pi}{\alpha}=f_{Z} \left( -\frac{4G_{F}}{\sqrt{2}}V^{*}_{ts}V_{tb}\right)&\\
&\Rightarrow f_{Z}=-\frac{1}{2\sqrt{2}G_{F}V_{tb}V^{*}_{ts}}\frac{4\pi}{\alpha}\frac{1}{8}\frac{g^{2}}{M_{Z}^{2}}V^{*}_{4k}V_{4l}&
\end{aligned}
\end{equation}
\section{Wilson Coefficients and LFUV}
\begin{figure}
\begin{center}
\includegraphics[width=3in]{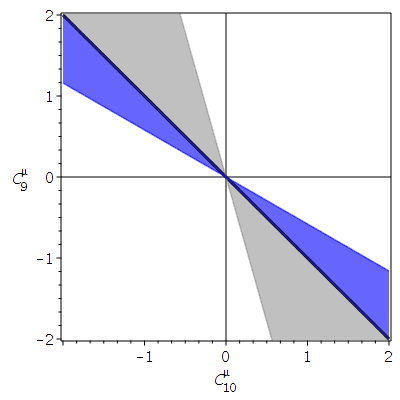}
\end{center}
\caption{\label{Fig.1} Allowed regions for the Wilson coefficients for both cases B (wedge in gray) and A (wedge in blue) for \(\beta=-1/\sqrt{3}\) \cite{main article} compared to \(\beta=1/\sqrt{3}\) (thick black line)}
\end{figure}

The NP contributions from \(Z_{\mu}\) and \(Z_{\mu}^{'}\) to the Wilson coefficients can be written in terms of the quantities \(f_{Z^{'}}\) and \(f_{Z}\) as
\begin{equation}
C_{9}^{ij}=f_{Z^{'}}\left( -\lambda_{ij}+\frac{1+3\tan^{2}\theta_{W}}{2}\right) \delta_{ij}+f_{Z}\left( -1+3\tan^{2}\theta_{W}\right) \delta_{ij}
\end{equation}
and
\begin{equation}
C_{10}^{ij}=f_{Z^{'}}\left( \lambda_{ij}+\dfrac{\tan^{2}\theta_{W}-1}{2}\right) \delta_{ij}+f_{Z}\left( 1+\tan^{2}\theta_{W}\right) \delta_{ij}
\end{equation}
where \(\lambda_{ij}=V^{*}_{1j}V_{1j}\).
Even though our model allows for lepton flavor violating transition with different leptons in the final state \((i\neq j)\), these processes have not been observed up to now, so, assuming that they are suppressed, we set their coefficients to zero. The solution \(f_{Z^{'}}=0\), i. e. the NP contribution is zero, meaning that there would be no LFUV, will be discarded. So, we are left with \(\lambda_{ij}=0\) for \(i\neq j\).
By definition\\
\begin{equation}
\label{lambdaij}
\lambda_{ij}=0\Longrightarrow V^{*}_{1i}V_{1j}=0\\
\end{equation}
Equation \eqref{lambdaij} does not necessarily imply that both \(V\) matrix elements have to be zero; one rotation matrix entry can be non-zero for a generation \(i\) (e.g. \(i=1\)) while the other two entries (e.g. \(j=2,3\)) are zero, ensuring that the above annihilation is realized. We denote with \(I\) the generation for which the entry for the rotation matrix is non-zero, and with \(i\) the other generations. We get
\begin{equation}
\label{CI}
\begin{aligned}
C_{9}^{I}=f_{Z^{'}}\left( -\lambda_{I}+\frac{1+3\tan^{2}\theta_{W}}{2}\right) +f_{Z}\left( -1+3\tan^{2}\theta_{W}\right)\\
C_{10}^{I}=f_{Z^{'}}\left( \lambda_{I}+\frac{\tan^{2}\theta_{W}-1}{2}\right) +f_{Z}\left( 1+\tan^{2}\theta_{W}\right)
\end{aligned}
\end{equation}
and
\begin{equation}
\label{Ci}
\begin{aligned}
C_{9}^{i}=f_{Z^{'}}\left(\frac{1+3\tan^{2}\theta_{W}}{2}\right) +f_{Z}\left( -1+3\tan^{2}\theta_{W}\right)\\
C_{10}^{i}=f_{Z^{'}}\left( \frac{\tan^{2}\theta_{W}-1}{2}\right) +f_{Z}\left( 1+\tan^{2}\theta_{W}\right)
\end{aligned}
\end{equation}
Inverting relations \eqref{Ci} we get
\begin{equation}
\label{Z' contribution}
f_{Z^{'}}=\frac{1+\tan^{2}\theta_{W}}{4\tan^{2}\theta_{W}}C_{9}^{i}-\frac{-1+3\tan^{2}\theta_{W}}{4\tan^{2}\theta_{W}}C_{10}^{i}
\end{equation}
From the system of equations \eqref{Ci} and \eqref{CI} we get
\begin{equation}
\label{relation between Ci, CI and Z' contribution}
2\lambda_{I}f_{Z^{'}}=C_{10}^{I}-C_{9}^{I}-C_{10}^{i}+C_{9}^{i}
\end{equation}
We now have to identify which index corresponds to which lepton, knowing that, based on phenomenological constraints, the electronic NP contribution to the effective Hamiltonian \(C_{9,10}^e\) is absent. 
\begin{itemize}
\item[(i)]
If we identify the electron with the index \(i\) (for which the entry for the rotation matrix vanishes), we set \(C_{9,10}^{i}=0\). Equation \eqref{Z' contribution} implies that \(f_{Z^{'}}=0\), solution that has to be discarded.
\item[(ii)] If the electron is identified with the index \(I\), the coefficients \(C_{9,10}^{I}\) must be set to zero in this case, and the remaining index \(i\) would correspond to the other two generations. In this case, eq. \eqref{relation between Ci, CI and Z' contribution} yields costraints on the non-vanishing NP Wilson coefficients for \(\mu\) and \(\tau\)
\begin{equation}
\label{final result}
\frac{C_{9}^{\mu}}{C_{10}^{\mu}}=\frac{2\tan^{2}\theta_{W}+\lambda_{e}\left( 1-3\tan^{2}\theta_{W}\right) }{2\tan^{2}\theta_{W}-\lambda_{e}\left( 1+\tan^{2}\theta_{W}\right) }
\end{equation}
\end{itemize}
Due to the unitarity of the \(7\times7\) rotation matrix \(V\), we have
\begin{equation}
\sum_{i=1}^{7} \mid V_{1i} \mid^{2}=1\Longrightarrow \lambda_{I}=\mid V_{1I} \mid^{2}=1-\sum_{i=4}^{7} \mid V_{1i} \mid^{2}
\end{equation}
which means that \(\lambda_{e}\in \left[0,1\right]\). For \(\lambda_{e}=1\), equation \eqref{final result} gives the exact equality \(C_{9}^{\mu}/C_{10}^{\mu}=-1\) \(\left( \sin^{2}\theta_{W}\simeq0.235\right)\). As \(\lambda_{e}\) decreses, the value of \(C_{9}^{\mu}/C_{10}^{\mu}\) increses until we get the exact equality \(C_{9}^{\mu}/C_{10}^{\mu}=1\) for \(\lambda_{e}=0\). Thus, the one-dimentional scenario of the global analysis that favors NP in \(C_{9}^{\mu}=-C_{10}^{\mu}\) is allowed in our model only for \(\lambda_{e}=1\). As for the other two scenorios (NP in \(C_{9}^{\mu}=-C_{9}^{\mu}\) and NP in \(C_{9}^{\mu}\) \cite{global analysis}), they cannot be described in the framework of our model, since no FCNC arise for right-handed quarks because their interaction terms are diagonal in flavor space (\eqref{lagZ} and \eqref{lagZ'}). Thus \(C_{9'}^{\mu}=0\).\\
Figure \eqref{Fig.1} shows that the allowed regions for the Wilson coefficients in both, case A and B \cite{main article} for \(\beta=-1/\sqrt{3}\) agree with our case \(\beta=-1/\sqrt{3}\) only for \(\lambda_{e}=1\). \\
In summary, not only the electron (first generation of SM leptons) has to be identified with the non-vanishing entry in the rotation matrix \(V\), but it has to be also a mass eigenstate in order to have non-vanishing NP contributions to Wilson coefficients, for both \(\mu\) and \(\tau\) that agree with the favored one-dimentional scenario of NP in \(C_{9}^{\mu}=-C_{10}^{\mu}\). 
\section{Conclusion}
In an attempt to give an explanation to the deviations from the Standard Model in \(b\rightarrow s l^+ l^-\) transitions, we have investigated a non-minimal version of the 331 models. In order to explain LFUV, five lepton triplets are required in this set, where the additional heavy gauge bosons and fermions have electric charges similar to those of the SM particles. We have shown how this model could explain these experimentally observed deviations, provided that the latter are dominated by neutral gauge bosons \(Z_{\mu}\) and \(Z^{'}_{\mu}\) contributions. Due to this assumption, the model turns out to have no right-handed currents, thus, not able to accomodate NP contribution in  \(C_{9}^{\mu}=-C_{9}^{\mu}\). When constraints on the mixing matrices between interaction and mass fermion eigenstates are put in light of the absence of contributions to \(b\rightarrow s e^+ e^-\), our model becomes able to accomodate significant NP contribution in \(C_{9}^{\mu}=-C_{10}^{\mu}\), in agreement with an NP scenario favored by global fits.
\ack
We are very grateful to the Algerien Ministry of Higher Education and Scientific Research and the DGRSDT for their finantial support.

\end{document}